\RequirePackage{ifpdf}
\ifpdf 
\documentclass[pdftex]{sigma}
\else
\documentclass{sigma}
\fi

\begin{document}
\allowdisplaybreaks

\renewcommand{\PaperNumber}{035}

\FirstPageHeading

\ShortArticleName{Calogero Model(s) and Deformed Oscillators}

\ArticleName{Calogero Model(s) and Deformed Oscillators}

\Author{Marijan MILEKOVI\'C $^\dag$, Stjepan MELJANAC $^\ddag$ and
Andjelo  SAMSAROV $^\ddag$}

\AuthorNameForHeading{M. Milekovi\'c,  S. Meljanac and A.
Samsarov} \Address{$^\dag$~Physics Department, Faculty of Science,
Bijeni\v cka c.~32, 10002 Zagreb, Croatia }
\EmailD{\href{mailto:marijan@phy.hr}{marijan@phy.hr}}
\Address{$^\ddag$~Rudjer Bo\v skovi\'c Institute, Bijeni\v cka c.
54, 10002 Zagreb, Croatia }
\EmailD{\href{mailto:meljanac@irb.hr}{meljanac@irb.hr}}

\ArticleDates{Received November 30, 2005, in f\/inal form March
02, 2006; Published online March 17, 2006}

\Abstract{We brief\/ly review some recent results concerning
algebraical (oscillator) aspects  of the $N$-body single-species
and multispecies Calogero models in one dimension. We  show how
these models emerge from the matrix generalization of the
harmonic oscillator Hamiltonian. We make some comments on the
solvability of these models.}


\Classification{81R12; 81R15; 81Q05; 46L65}

\section{Introduction}

There exists  a very limited number of exactly solvable many-body
systems, even in one dimension~(1D)~\cite{Mattis,Sutherland} and
the  Calogero model \cite{Calogero1,Calogero2} is surely one of
the most famous and exhaustively studied examples of such models.
In its original version, the Calogero model describes $N$
identical, spinless, nonrelativistic particles on the line which
interact through an inverse-square two-body interaction and are
subjected to a common conf\/ining harmonic force. Although being
1D problem, its solution is far from being straightforward and it
is really amazing how this  model and its various descedants,
known as Calogero--Sutherland--Moser systems~\cite{Vinet}, have so
deep and profound impact on various branches of  physics  and
mathematics, ranging from condensed matter systems and black hole
physics~\cite{Kawakami, Gibbons} to random matrices and
strings~\cite{Sutherland1,Kapustin}. Why is this  so and what
motivates physicist and mathematicians to study such a class of
models? Maybe the best answer, more that three decades ago, was
of\/fered by Calogero himself in the opening section of his
seminal paper~\cite{Calogero2}. In his own words ``{\it A
motivation is perceived in the insight that exact solutions, even
of oversimplified models, may provide and in the possibility to
assess the reliability of approximation techniques that can be
used in more realistic contexts, by first testing them in exactly
solvable cases. Moreover, for some physical problems a
$1$-dimensional schematization may indeed be appropriate.}''

The single-species $N$-body Calogero model  is def\/ined by the
following Hamiltonian ($i,j = 1,2,\ldots,N$):
\begin{gather}\label{eq1}
H=-\frac{\hbar^2}{2m} \sum_{i} \frac{\partial^2}{\partial x_i^2} +
\frac{m\omega^2}{2}\sum_{i} x_i^2 + \frac{\hbar^2 \nu (\nu -
1)}{2m} \sum_{i\neq j }\frac{1}{(x_i-x_j)^2},
\end{gather}
where $\nu (\nu - 1)\equiv g$ is the dimensionless coupling
constant, equal for  all particles. The well-known stability
condition requires $g \geq -1/4$. The constant $m$ is the mass of
particles and $\omega$ is the strength of a common harmonic
conf\/inement potential. The ground state of the
Hamiltonian~\eqref{eq1}
 is known to be of the  highly correlated, Jastrow form
\begin{gather}\label{eq2}
\Psi_0 (x_1,x_2,\ldots, x_N)=\left( \prod_{i<j} |x_i -x_j|^{\nu}
\right)e^ {-\frac{m\omega}{2\hbar} \sum_{i}  x_i^2} \equiv \Delta
\, e^{-\frac{m\omega}{2\hbar}\sum_{i}  x_i^2}.
\end{gather}
The  corresponding ground state energy  depends on $\nu $
explicitly and is given by
\begin{gather}\label{eq3}
E_0=\omega \left(\frac{N}{2} + \frac{\nu N (N-1)}{2}\right).
\end{gather}
As conjectured by Calogero \cite{Calogero1}, the remainder of the
eigenspectrum of the model \eqref{eq1}
 coincides (except for a constant shift of all energy levels)
 with the energy spectrum of the system of free harmonic oscillators.
 An explicit mapping of the model \eqref{eq1} to the set of $N$ free  harmonic oscillators
 was found quite recently in~\cite{Gurappa} and later applied to the other Calogero-like systems~\cite{Khare}.

One can solve this model exactly and f\/ind out the complete set
of energy eigenvalues \mbox{either} by following the traditional
approach~\cite{Calogero1, Calogero2, Sutherland2} (see also the
recent paper by Hallnas and Langmann~\cite{Langmann}, who
brief\/ly review this and some other approaches, including their
own new solution algorithm for solving \eqref{eq1})  or by
employing its underlying $S_N$ (permutational) algebraic
structure~\cite{Brink}. The later  approach, based on the
particular set of creation and annihilation operators, is
considerably simpler than the original one \cite{Calogero1,
Calogero2} and yields an explicit expression for the wavefunctions
in terms of action of creation operators on the ground state
(vacuum). We review this approach in Section~2.

The single-species Hamiltonian \eqref{eq1} can be easily
generalized to the multispecies variant of the model
\cite{Meljanac1, Melj} and even to the arbitrary  dimensions
\cite{Meljanac2}. The   multispecies Calogero model, as a
generalization of the Calogero model \eqref{eq1} to the system of
distinguishable particles, is worth to study for (at least) two
reasons. It exhibits some novel features not encountered in the
model with identical particles and represents a playground for the
testing  mathematical tools, developed earlier for the
single-species Calogero model (see comments below).  It is  also
related to the notion of the  Haldane exclusion statistics
\cite{Haldane} and the origin of this relation can be traced back
to the model~\eqref{eq1}. Namely, the Calogero model \eqref{eq1}
provides a microscopic realization of the generalized Haldane
exclusion statistics, where the role of the Haldane statistical
parameter is played by the  coupling constant $g$ in the
inverse-square interaction \cite{BW,MS}. In Haldane's formulation
of statistics there is a possibility of having particles of
dif\/ferent species,  with a~statistical  parameter depending on
the  species coupled. Thus, the  extension of the model
\eqref{eq1} to the multispecies case is a natural step \cite
{Ouvry, Sen, Mashkevic}. We also mention that the simplest
multispecies model, with just two dif\/ferent kind of particles
\cite{Melj}, has been recently reconsidered in the supersymmetric
framework and it has been claimed that the corresponding
Hamiltonian  might describe the motion of electrons in a
one-dimensional superconductor~\cite{Guhr}.

Distinguishability of the species can be introduced by allowing
particles to have dif\/ferent masses ($m \rightarrow m_{i}$) and
dif\/ferent couplings ($\nu \rightarrow \nu_{ij}$) to each other.
(In fact, this  generalization has already been suggested in
\cite{Calogero2},
 although not using statistical arguments.)
The common feature of these generalizations (including higher
dimensional cases) is a new long-range three-body interaction that
appears in the Hamiltonian. Its character and strength is
determined by the parameters of the two-body interaction. In 1D
multispecies variant of~\eqref{eq1}, this three-body interaction
can be switched of\/f by a suitable chosen connection between
coupling constants $\nu_{ij}$ and masses $m_{i}$ of the particles
but for the models in $D>1$  this interaction is generic and
cannot be turned of\/f. Contrary to the 1D single-species model
\eqref{eq1}, in all above cases only a limited set of exact
eigenstates and eigenenergies are known and the complete solution
of the multispecies problem, even in 1D, is still lacking. It is
interesting that  the $S_N$-based approach (in a slightly
generalized form), so successfully  used in \cite{Brink} to solve
Hamiltonian \eqref{eq1}, does not hold  for the 1D multispecies
model. Instead of $S_N$ structure, one can try to employ
underlying SU(1,1) structure of the multispecies Hamiltonian and
def\/ine an another set of creation and annihilation operators
(construct from SU(1,1) generators) but, as we explain  in
Section~4, this approach also  has limitations.

Section 3 is an interlude. We show how single- and multispecies 1D
Calogero models emerge from the matrix generalization of the
single harmonic oscillator Hamiltonian \cite{Meljanac3}, i.e.\ we
adopt a view that the both models represent, in a some sense, a
``deformed'' harmonic oscillator, ``deformation'' being encoded in
an unusual commutation relations. Section~5 is a concluding
section.

\section[Single-species Calogero model in 1D and $S_N$-extended Heisenberg algebra]{Single-species
Calogero model in 1D\\ and $\boldsymbol{S_N}$-extended Heisenberg
algebra}

In this section we brief\/ly review an operator solution to the
Calogero model \eqref{eq1}, based on $S_N$-extended Heisenberg
algebra, and discuss Fock space of this algebra. We closely follow
references~\cite{Brink, Meljanac4, Meljanac5}.

In order to simplify the analysis, we extract Jastrow factor
$\Delta $ from the ground state $\Psi_{0}(x_1,x_2$, $\ldots, x_N)$
and def\/ine  new vacuum $|\tilde{0}\rangle = \tilde{\Psi}_{0} =
\Delta^{-1} \Psi_{0}$. This generates a similarity transformation
on  Hamiltonian \eqref{eq1}, which leads to an another Hamiltonian
$\tilde{H}$ i.e.\ $\tilde{H} = \Delta^{-1}H \Delta $.

We f\/ind $\tilde{H}$ as
\begin{gather}\label{eq4}
\tilde{H} = -\frac{1}{2m}\sum_{i=1}^{N}\frac{{\partial}^{2}}
{\partial x_{i}^{2}} + \frac{{m\omega}^{2}}{2}\sum_{i=1}^{N }
x_{i}^{2} - \frac{1}{2m} \sum_{i \neq j }\frac{\nu}
{(x_{i}-x_{j})}\left( \frac{\partial}{{\partial} x_{i}}
 -  \frac{\partial}{{\partial} x_{j}}\right).
\end{gather}
Instead of the Hamiltonian \eqref{eq1}, it is then convenient to
work with the transformed Hamiltonian~$\tilde{H}$. We def\/ine a
covariant derivative $D_i$ \cite{Brink, Polychronakos} by
\begin{gather}\label{eq5}
D_i = \partial_i + \nu \sum_{j\neq i} \frac{1}{(x_i - x_j)}\, (1 -
K_{ij}),
\end{gather}
that close under the commutation,  $[D_i,D_j]=0$. Here
$K_{ij}$'s are  the generating elements of the  symmetric group
$S_N$
\begin{gather*}
K_{ij}=K_{ji}, \qquad (K_{ij})^2=1, \qquad
K_{ij}^{\dagger}=K_{ij},
\\
K_{ij}K_{jl}=K_{jl}K_{il}=K_{il}K_{ij}, \qquad i\neq j, \quad
i\neq l,\quad j\neq l ,
\end{gather*}
which interchange the particles with labels $i$ and $j$, for
example  $K_{ij}x_j=x_iK_{ij}$.

In terms of covariant derivatives and coordinates one can def\/ine
ladder operators $a_i$ and $a_i^{\dagger}$ as (we put all
unimportant constants equal to one):
\begin{gather}\label{eq6}
a_i^{\dagger}=\frac{1}{\sqrt 2} (x_i - D_i), \qquad
a_i=\frac{1}{\sqrt 2} (x_i + D_i).
\end{gather}
The physical space of $N$ identical particles is represented
either by symmetric or antisymmetric wave functions. It can be
shown that all necessary information is actually contained in the
totally symmetric space. Restriction to the totally symmetric
space  yields Hamiltonian
\begin{gather*}
\tilde H =\Delta ^{-1}\, H \,\Delta = \frac{1}{2}\sum_{i}
\{a_i,a_i^{\dagger}\}=\sum_{i}a_i^{\dagger}a_i+E_0=
\frac{1}{2}\sum_{i} \big(X_i^2 + P_i^2\big),
\end{gather*}
where $P_i= -\imath D_i$ and $X_i=x_i$. In deriving these
expressions, we used the commutation rules that follow directly
from the def\/initions \eqref{eq5} and \eqref{eq6}:
\begin{gather}
[X_i,P_i]=\imath \, \delta_{ij}\left(1 + \nu \sum_{k=1}^N
K_{ik}\right) - \imath \nu K_{ij},
\nonumber\\
[a_i,a_j^{\dagger}]= \delta_{ij}\left(1+ \nu \sum_{k=1}^N
K_{ik}\right) - \nu K_{ij}, \qquad [a_i^{\dagger},a_j^{\dagger}]
=[a_i,a_j]=0 ,\label{eq8}
\end{gather}
together with the vacuum  conditions $a_i|\tilde{0}\rangle =0$,
$K_{ij}|\tilde{0}\rangle= +|\tilde{0}\rangle $.

The algebra of creation and annihilation operators \eqref{eq8} is
known as $S_N$-extended Heisenberg algebra. The physical Fock
space of $\tilde H $ is spanned by $S_N$-symmetric states $(
\prod_{n_k}{\cal {B}} _k^{\dagger n_k} |\tilde{0}\rangle )$,
\[
\tilde H \left( \prod_{n_k}{\cal {B}}_k^{\dagger n_k}
|\tilde{0}\rangle \right)= \left[ E_0 + \sum_{k=1}^N k\,n_k
\right] \left( \prod_{n_k}{\cal {B}} _k^{\dagger n_k}
|\tilde{0}\rangle\right) ,
\]
where ${\cal {B}} _k=\sum_{i} a_i^k$ are collective,
$S_N$-symmetric operators. For further details about algebra of
these operators and Fock space structure we refer the reader to
references~\cite{Jonke1,Jonke2}.

As it stands, the algebra \eqref{eq8} is consistent for all values
of the parameter $\nu$. The only additional restriction on the
$\nu$ may come from the representation of the algebra on the Fock
space. For example, the states in the complete Fock space of
algebra \eqref{eq8} should have positive norm. Apart from its
particular realization, $S_N$-invariant Heisenberg algebra is
basically a multimode oscillator algebra with permutational
invariance. We developed earlier a gene\-ral techniques for
analyzing such a class of algebras \cite{Meljanac5, Meljanac6}.
Information about structure of the Fock space  is encoded  either
in (i) action of annihilation operators on monomial states in Fock
space, $a_i\, (a_1^{\dagger n_1}\cdots a_N^{\dagger
n_N}|\tilde{0}\rangle )$ or in (ii) Gram matrices of scalar
products in Fock space, $A_{i_N\cdots i_1;j_1 \cdots j_N}= \langle
\tilde{0}|a_{i_N}\cdots a_{i_1}a_{j_1}^{\dagger}\cdots
a_{j_N}^{\dagger}|\tilde{0} \rangle $. For the algebra \eqref{eq8}
there is no closed form available
 for the approach (i) but one can use (slightly cumbersome) recursion
 relations for it \cite{Meljanac5}, which may help   to obtain matrix elements of the Gram matrix (ii).

The Gram matrix for the one-particle  states
$a_i^{\dagger}|\tilde{0}  \rangle $ is of order $N$ and has only
two distinct entries. The  Gram matrix for two-particle  states
$a_i^{\dagger}a_j^{\dagger}|\tilde{0}  \rangle $ is of order $N^2$
and has four distinct entries and so on. For the convenience, we
display below the Gram matrix of one-particle states, i.e.
\[
\left(\begin {array}{cccc}
1+\nu (N-1) & -\nu & \cdots & -\nu \\
-\nu & 1+\nu (N-1) & \cdots & -\nu \\
\cdots & \cdots & \cdots & \cdots  \\
-\nu & -\nu &  \cdots & 1+\nu (N-1)
\end{array}
\right).
\]
Its  eigenvalues and eigenvectors are
\[
\begin{tabular}{|c|c|c|c|} \hline
Eigenvalue & Degeneracy & Eigenvector &  Comments \\ \hline \hline
1 & 1 & ${\cal {B}}_1 ^{\dagger}|\tilde{0} \rangle $&  \\ \hline
$1+ N \nu $ &$ N-1$ &$ ( a_1^{\dagger} - a_i^{\dagger} )|\tilde{0}
\rangle$ &$ i\neq 1$\\ \hline \hline
\end{tabular}
\]
This Gram matrix is positive def\/inite, i.e.\ there are no
eigenvectors with negative norm if $\nu > -\frac{1}{N}$. At $\nu =
-\frac{1}{N}$, there is a critical point where the ($N-1$) states
$( a_1^{\dagger} - a_i^{\dagger} )|0\rangle$ have zero  norm. One
can show that for two-, three-, and many-particle states the same
condition for positivity of eigenvalues is required
\cite{Meljanac4, Meljanac5}. There exists the universal critical
point at $\nu = -\frac{1}{N}$, where all matrix elements of the
arbitrary $k$-multi-state Gram matrix are equal
 to~$\frac{k!}{N^k}$. In the limit $\nu N \rightarrow -1$ the system of deformed oscillators \eqref{eq8}
exhibits singular behaviour. There survives only one oscillator,
of the type $ {\cal {B}}_1 ^{\dagger}$, describing the motion of
the center-of-mass coordinate. All other ($N-1$) relative
coordinates collapse into the same point-the center of mass. This
means that the relative coordinates, the relative momenta and the
relative energy are all zero at the critical value of the
parameter $\nu $. This can be also verif\/ied independently by
applying large-$N$ collective f\/ield theory to the Hamiltonian
$\tilde H$ \cite{Meljanac4}. However, one should bear in mind that
the interval $\nu \in (-\frac{1}{N},0)$ is physically acceptable
for the Hamiltonian $\tilde H$, describing $N$ oscillators with
$S_N$-extended Heisenberg algebra, but it is not allowed for the
original Calogero Hamiltonian \eqref{eq1} since the wave
functions, containing Jastrow factor, diverge at the coincidence
point for negative values of~$\nu $. This reminds the fact that
the two Hamiltonians are not unitary (i.e.\ physically)
equivalent.

\section{Interlude: matrix quantum deformed oscillators\\ and
emergence of the Calogero-like models}

In the preceding section we have seen that the Hamiltonian $\tilde
H$ was still given as quadratic form in coordinates $X_i=x_i$  and
(generalized) momenta $P_i= -\imath D_i$, i.e.\ in the form of the
harmonic oscillator Hamiltonian. It is then tempting to start with
the most general harmonic oscillator Hamiltonian and try to
reproduce Calogero-like Hamiltonian(s). In this section we
present, following~\cite{Meljanac3}, such construction. The
f\/inal goal is to obtain, by adjusting parameters, not only a
single-species Calogero model \eqref{eq1} but also a multispecies
one.

Let us consider a matrix generalization of harmonic oscillator
Hamiltonian ($\omega=1$)
\begin{gather}\label{eq9}
{\boldsymbol {H}} =\frac{1}{2}\big({\boldsymbol
{P}}M^{-1}{\boldsymbol {P}} +
 {\boldsymbol {X}}M{\boldsymbol {X}}\big).
\end{gather}
where $M_{ij}= m_i \delta_{ij}$ is non-singular mass matrix
($m_i>0$) and $\boldsymbol{P}=-\imath \boldsymbol{D}$  and
$\boldsymbol{X}$ are generalized momentum  and coordinate
matrices respectively, with operator-valued matrix elements:
\[
{\boldsymbol{P}}=\left(\begin{array}{cccc}
P_{11} & P_{12} & \cdots & P_{1N} \\
P_{21} & P_{22} & \cdots & P_{2N} \\
\cdots & \cdots & \cdots & \cdots  \\
P_{N1} & P_{N2} &  \cdots & P_{NN}
\end{array}
\right) ,\qquad {\boldsymbol{X}}=\left(\begin{array}{cccc}
x_{1} & 0 & \cdots &0 \\
0 & x_{2} & \cdots & 0 \\
\cdots & \cdots & \cdots & \cdots  \\
0 & 0 &  \cdots & x_{N}
\end{array}
\right).
\]
We assume the following deformed matrix commutation relations:
\begin{gather}\label{eq10}
[{\boldsymbol{X}},{\boldsymbol{P}}]= i {\boldsymbol{\Lambda}}
\quad \Rightarrow \quad [{\boldsymbol {X}},{\boldsymbol{D}}]= -
{\boldsymbol{\Lambda}},
\end{gather}
where $\boldsymbol{\Lambda}$ is a Hermitian matrix with constant,
real and symmetric matrix elements
\[
{\boldsymbol{\Lambda}}=\left(\begin{array}{cccc}
1 & 0 & \cdots &0 \\
0 & 1 & \cdots & 0 \\
\cdots & \cdots & \cdots & \cdots  \\
0 & 0 &  \cdots & 1
\end{array}
\right)  + \left(\begin{array}{cccc}
0 & \nu_{12} & \cdots & \nu_{1N} \\
\nu_{12} & 0 & \cdots & \nu_{2N} \\
\cdots & \cdots & \cdots & \cdots  \\
\nu_{1N} & \nu_{2N} &  \cdots & 0
\end{array}
\right).
\]
In terms of matrix elements, the commutation relations
\eqref{eq10} read
\begin{gather}\label{eq11}
D_{ij}x_j - x_i D_{ij}=\nu_{ij}, \qquad D_{ii}x_i - x_i D_{ii}=1.
\end{gather}
There are many solutions of these equations, since adding a
coordinate, i.e.\ $\boldsymbol{X}$-dependent, part to
$\boldsymbol{D}$ does not af\/fect them. We can always transform
$\boldsymbol{D}$ by an arbitrary function $F({\boldsymbol{X}})$,
i.e.\ ${\boldsymbol{D}} \rightarrow  F({\boldsymbol{X}})^{-1}\,
{\boldsymbol{D}} \, F({\boldsymbol{X}})$. On the level of
Hamiltonian this generates a non-unitary  transformation
${\boldsymbol H}\rightarrow
F({\boldsymbol{X}})^{-1}\,{\boldsymbol H}\, F({\boldsymbol{X}})$.

We restrict ourselves to transformations def\/ined by
\[
F({\boldsymbol {X}}) = \prod_{i<j}(x_i - x_j)^{\theta_{ij}} , \qquad
\theta_{ij}= \theta_{ji}.
\]
A corresponding class of solutions of equations \eqref{eq11} is
given by
\[
D_{ij}= \delta_{ij} \left(\frac{\partial}{\partial x_i} +
\sum_{k\neq i}\theta_{ik} \frac{1}{(x_i-x_k)}\; \right ) -
\nu_{ij} \frac{(1-\delta_{ij})}{(x_i-x_j)}.
\]
To summarize: we ended up with a class of (matrix) Hamiltonians
that depends on three parameters, namely masses $m_i$,
``deformation'' parameters $\nu_{ij}=\nu_{ji}$ and  ``gauge''
parameters $\theta_{ij}=\theta_{ji}$. To obtain a one-dimensional
models from \eqref{eq9}, one has to perform a suitable
``reduction''. The following ``reduction'' was proposed in
\cite{Meljanac3}:
\begin{gather}\label{eq12}
H= {\rm Tr}\, ({\boldsymbol{H}} \cdot{\boldsymbol{J}}),
\end{gather}
where the constant matrix ${\boldsymbol{J}}$ has the form
\[
{\boldsymbol{J}}=\left(\begin{array}{cccc}
1 & 1 &\cdots &1  \\
1 & 1  &  \cdots &1      \\
\cdots & \cdots & \cdots & \cdots \\
1 & 1  &\cdots & 1
\end{array}
\right).
\]
The ``reduction'' procedure outlined here dif\/fers from that
followed in~\cite{Gorsky} in the way how Calogero model appears.
The Hamiltonian obtained here is not just the trace of the matrix
Hamiltonian ${\boldsymbol H}$, as in \cite{Gorsky}, but is given
by the trace over $\boldsymbol{H}\cdot \boldsymbol{J}$.

One can easy convince himself that the prescription \eqref{eq12}
and the following set of parameters  $\{ m_i=m; \nu_{ij}=\nu;
\theta_{ij}=0 \}$ leads to Calogero Hamiltonian \eqref{eq1}.
Similarly, by choosing $\{ m_i=m; \nu_{ij}=\nu; \theta_{ij}=
\nu_{ij}=\nu\}$ one reproduces Hamiltonian \eqref{eq4}.

More interesting choice is $\{ m_i; \nu_{ij}=\nu_{ji};
\theta_{ij}=0 \}$, which results in Hamiltonian of {\it
multispecies} 1D Calogero model, with three-body
interaction~\cite{Meljanac1}:
\begin{gather}
H=-\frac{1}{2}\sum_i \frac{1}{m_i}\frac{\partial^2}{\partial
x_i^2}
 +\frac{1}{2}\sum_i m_ix_i^2 +\frac{1}{4}\sum_{i\neq j }
 \frac{\nu_{ij}(\nu_{ij}-1)}{(x_i-x_j)^2} \left(\frac{1}{m_i}+\frac{1}{m_j}\right)\nonumber\\
\phantom{H=}{} +{\underbrace {\frac{1}{2}\sum_{i,j,k
\neq}(\frac{\nu_{ij}\nu_{jk}}{m_j})\frac{1}{(x_j-x_i)
(x_j-x_k)}}_{\rm {3-body\; interaction}}}.\label{eq13}
\end{gather}
The {\it multispecies} analogue of Hamiltonian $\tilde {H}$ (4),
i.e.\ ``gauge'' transformed Hamiltonian
 $ {\tilde {\boldsymbol H}}= F({\boldsymbol {X}})^{-1}\,{\boldsymbol H}\, F({\boldsymbol {X}})$,
  is reproduced by choosing
$\{ m_i; \nu_{ij}=\nu_{ji}; \theta_{ij}= \nu_{ij}\}$, and reads
\begin{gather}\label{eq14}
{\tilde {H}}=-\frac{1}{2}\sum_i
\frac{1}{m_i}\frac{\partial^2}{\partial x_i^2} +\frac{1}{2}\sum_i
m_ix_i^2 - \frac{1}{2} \sum_{i \neq j }\frac{{\nu}_{ij}}
{(x_{i}-x_{j})}\left(\frac{1}{m_{i}} \frac{\partial}{{\partial}
x_{i}}
 - \frac{1}{m_{j}} \frac{\partial}{{\partial} x_{j}}\right).
\end{gather}
A few additional remarks concerning the Hamiltonian \eqref{eq13}
are in order.

First,  it describes distinguishable particles on the line,
interacting with harmonic, two-body and three-body potentials. Its
ground state  reads
\begin{gather}
|0 \rangle \equiv \Psi_0 (x_1,x_2,\ldots, x_N)=\prod_{i<j} |x_i
-x_j|^{\nu_{ij}}\; e^ {-\frac{1}{2}
\sum_{i} m_i x_i^2}=\Delta \; e^ {-\frac{1}{2}\sum_{i} m_i x_i^2} ,\nonumber\\
E_0= \frac{N}{2}+ \frac{1}{2}\sum_{i\neq j}\nu_{ij}.\label{eq15}
\end{gather}
and tends to the ground state of the single-species Calogero model
\eqref{eq2}, \eqref{eq3} in the limit \mbox{$\nu_{ij}\rightarrow
\nu$}, $m_i\rightarrow m$. The appearance of the generalized
Jastrow factor $\Delta= \prod\limits_{i<j} |x_i -x_j|^{\nu_{ij}}$
in \eqref{eq13} has the same origin as in the Calogero model
\eqref{eq1}. Namely, because of the singular nature of the
Hamiltonian for $x_i =x_j$, the wave function ought to have a
prefactor that vanish for the coincident particles. Also, the
stability condition demands that the two-body couplings
$g_{ij}=\nu_{ij}(\nu_{ij}-1)$ should be greater than
$-\frac{1}{4}$.

Let us consider the last term in \eqref{eq13} i.e.\ the three-body
interaction, more closely. If we put  $m_j=m={\rm const}$ in
\eqref{eq13}, symmetrize under the cyclic exchange of the indices
($i\rightarrow j \rightarrow k \rightarrow i $) and reduce the sum
to a common denominator using the identity
\[
\sum_{\rm cycl} \frac{1}{(x_i-x_j)(x_i-x_k)}=0,
\]
we obtain that the necessary condition for vanishing of the
three-body interaction is $\nu_{ij}=\nu$.  In this way, the
problem (13) is reduced to the ordinary $N$-body Calogero model
with harmonic and two-body interactions only.

For the general $\nu_{ij}$ and $m_j$, it can be shown that  the
following conditions for the absence of the three-body term must
hold~\cite{Meljanac1}:
\[
\frac{\nu_{ij}\nu_{jk}}{m_j}=\frac{\nu_{jk}\nu_{ki}}{m_k}=\frac{\nu_{ki}\nu_{ij}}{m_i}
\qquad \forall\; (i,j,k).
\]
The unique solution of these conditions   is $\nu_{ij}=\lambda \,
m_i\, m_j$,
 $\lambda $ being some universal constant. It should be mentioned that the above
arguments cannot be applied in higher dimensions, so the
three-body interaction is generic for the higher-dimensional
multispecies model \cite{ Meljanac2, Calogero3} and makes
corresponding Hamiltonian hard to solve.

We note in passing that this particular connection between masses
and interaction parameters was also displayed in~\cite{Sen,
Forrester}. In~\cite{Sen},
 the above conditions  arose from  the requirement that the asymptotic
 Bethe ansatz should be  applicable to the ground state of
 a multispecies many-body quantum system obeying mutual (exclusion) statistics,
 while in \cite{Forrester} its origin was not obvious.
 We of\/fer the simplest possible interpretation, namely
 these conditions eliminate the three-body interaction from the Hamiltonian.

\section[Multispecies Calogero model in 1D,
 $S_N$-extended Heisenberg algebra and $SU(1,1)$ algebra]{Multispecies Calogero model in 1D, \\
 $\boldsymbol{S_N}$-extended Heisenberg algebra and $\boldsymbol{SU(1,1)}$ algebra}

Now, we may  attempt to solve the Hamiltonian of the 1D
multispecies model \eqref{eq13}. Motivated by the successful
operator approach to single-species model \eqref{eq4} through
$S_N$-extended Heisenberg algebra, one can  try to apply the same
techniques to multispecies Hamiltonian. We shall discuss some
obstacles to this approach, f\/irst recognized in~\cite{Ouvry}.

Following the similar procedure as in Section 2, we def\/ine a
slightly generalized covariant derivatives $(i=1,2,\ldots,N)$
\begin{gather}\label{eq16}
D_i = \partial_i +  \sum_{j\neq i} \frac{\nu_{ij}}{(x_i - x_j)}\,
(1 - K_{ij})
\end{gather}
such that
\[
[D_i,x_j]= \delta_{ij}\left(1+  \sum_{k=1}^N \nu_{ik}K_{ik}\right)
- \nu_{ij} K_{ij}.
\]
Without losing  generality, and in order to  simplify
calculations, we  set all masses equal to one. Contrary to the
previous case the commutator $[D_i,D_j]$ does not vanish in
general, except on the totally symmetric states. In this
particular case, the Hamiltonian \eqref{eq14} reads $\tilde {H}  =
\frac{1}{2}\sum_{i}\{a_i,a_i^{\dagger}\}$, where creation and
annihilation operators are def\/ined as in \eqref{eq6} (with
covariant derivatives \eqref{eq5} replaced by \eqref{eq16}) and
their algebra, restricted to the totally symmetric subspace, is
\[
[a_i,a_j^{\dagger}]= \delta_{ij}\left(1+  \sum_{k=1}^N
\nu_{ik}\right) - \nu_{ij}, \qquad [a_i^{\dagger},a_j^{\dagger}]
=[a_i,a_j]=0.
\]
The vacuum state \eqref{eq15}, $|\tilde{0}\rangle = \Delta^{-1}
|0\rangle $, is annihilated by $a_i$. On this vacuum, we might try
to build the towers of excited states. It can be checked that the
center of mass excitations, ${\cal {B}} _1^{\dagger
n}|\tilde{0}\rangle =(\sum_{i} a_i^{\dagger })^n|\tilde{0}\rangle
$, are  indeed  eigenstates of $\tilde {H}$, with eigenenergies
$E=E_0 + n$. The same is true for the Calogero-like state ${\cal
{B}} _2^{\dagger}|\tilde{0}\rangle  =(\sum_{i} a_i^{\dagger
2})|\tilde{0}\rangle $. Its eigenenergy is $E=E_0 + 2$. However,
this procedure breaks down on the next Calogero-like state, namely
${\cal {B}} _3^{\dagger} |\tilde{0}\rangle =(\sum_{i} a_i^{\dagger
3})|\tilde{0}\rangle $. Calculation shows that
\[
\tilde {H}({\cal {B}} _3^{\dagger} |\tilde{0}\rangle )= (E_0+3)
({\cal {B}} _3^{\dagger}|\tilde{0}\rangle )+ {\rm const}\cdot
\sum_{j \neq i } \frac{{\nu}_{ij}}{(x_{i}-x_{j})} (\sum_{l \neq i
}\nu_{il} - \sum_{l \neq j }\nu_{jl}) | \tilde{0} \rangle ,
\]
i.e.\ this state is an eigenstate of $\tilde {H}$ only if the all
coupling constants $\nu_{ij}$ are equal.

The above result is not so surprising. The particles are no more
identical and restriction to totally symmetric space is a too
strong requirement. Most of the eigenstates do not have
a~symmetrical form. Nevertheless, this algebraic approach still
reproduces correctly some of the excited states. There is a claim
in the literature~\cite{Veigy}, based on numerical and
perturbative analysis of the 3-particle  model, that there is no
algebra of raising and lowering operators for the multispecies
model which gives its complete spectrum. At the moment, we cannot
prove or disproves this claim.

As was shown in \cite{Meljanac1}, the Hamiltonian \eqref{eq14}
 still possesses hidden SU(1,1) structure and one can hope that the
 additional  set of excited states can be determined using an algebraic
 approach based on  SU(1,1) algebra. Let us  def\/ine operators
\begin{gather*}
T_{-} = -\frac{1}{2}\sum_{i}\frac{1}{m_{i}}\frac{{\partial}^{2}}
{\partial x_{i}^{2}} - \frac{1}{2} \sum_{i \neq j
}\frac{{\nu}_{ij}} {(x_{i}-x_{j})}\left(\frac{1}{m_{i}}
\frac{\partial}{{\partial} x_{i}}
 - \frac{1}{m_{j}} \frac{\partial}{{\partial} x_{j}}\right),
\\
T_{+} = \frac{1}{2}\sum_{i}m_{i} x_{i}^{2} \qquad T_{0} =
\frac{1}{2} \left(\sum_{i} x_{i}\frac{\partial}{{\partial} x_{i}}+
E_{0}\right),
\end{gather*}
which satisfy SU(1,1) commutation relations $[T_{-},T_{+}]=2
T_{0}$, $ [T_{0},T_{\pm}]=\pm T_{\pm}$. It is  possible to
construct new set of ladder operators, out of SU(1,1) generators
($X$ denotes center-of-mass coordinate and $M$ is total mass of
particles)
\[
A_1^{\pm}=\frac{1}{\sqrt 2} \left( \sqrt{ M }X \mp \frac{1}{\sqrt{
M }}\frac{\partial }{\partial X}\right) ,\qquad
A_2^{\pm}=\frac{1}{2} (  T_+ +  T_- )\mp T_0
\]
with algebra
\begin{gather*}
\tilde {H}= [A_2^{-}, A_2^{+}],\qquad [\tilde {H},A_1^{\pm}] = \pm
A_1^{\pm},\qquad [\tilde {H},A_2^{\pm}] = \pm 2  A_2^{\pm},
\\
[A_1^{-}, A_1^{+}] = 1,\qquad  [A_1^{-},A_2^{+}]=A_1^{+}, \qquad
[A_1^{-},A_2^{-}]=[A_1^{+},A_2^{+}]=0.
\end{gather*}
The inf\/inite tower of Fock space-states  are built on the vacuum
$|\tilde {0}\rangle $ as ($ n_1,n_2 =0,1,\ldots $)
\begin{displaymath}
|n_1, n_2\rangle \propto A_1^{+ n_1} A_2^{+ n_2}|\tilde {0}\rangle
.
\end{displaymath}
Hamiltonian $\tilde {H}$ is diagonal in this basis, i.e.
\begin{gather*}
\tilde{H} \; |n_1, n_2\rangle = ( E_0 + ( n_1 + 2 n_2 )  )|n_1,
n_2\rangle.
\end{gather*}
The energy spectrum is linear in quantum numbers and highly
degenerate. It can be shown that the dynamical symmetry
responsible for this degeneracy is SU(2) \cite{Meljanac1}.

If we further def\/ine an operator $B_{2}^{\pm}=A_{2}^{\pm} -
\frac{1}{2}{A_{1}^{\pm}}^{2}$, then we can split the Hamiltonian
$\tilde {H}$ in the two parts:
\begin{gather}
\tilde{H}_{CM}=\frac{1}{2} \{A_{1}^{-},A_{1}^{+}\}_+ ,\quad
\tilde{H}_{R}=  [B_{2}^{-},B_{2}^{+}], \quad [\tilde{H}_{CM},
A_{1}^{\pm}]=\pm   A_{1}^{\pm},\quad [\tilde{H}_{R},B_{2}^{\pm}] =
\pm 2  B_{2}^{\pm},
\nonumber\\
\tilde{H}_{CM}|\tilde{0}\rangle_{CM}=\frac{1}{2}
|\tilde{0}\rangle_{CM},\qquad \tilde{H}_{R}|\tilde{0}\rangle_{R}=
(\frac{N-1}{2} +\frac{1}{2}\sum_{i\neq
j}\nu_{ij})|\tilde{0}\rangle_{R}. \label{eq18}
\end{gather}
It is clear now that the spectrum \eqref{eq18} consists of
center-of-mass mode (CM) and the mode that describes relative
motion (R). Furthermore, it is easy to detect the critical point
in the Fock space, i.e.\ the forbidden range of coupling constants
$\nu_{ij}$, by inspecting the zero-norm states
\[
||B_{2}^{+}|\tilde{0}\rangle_{R}|| = 0,  \qquad _R\langle
\tilde{0} |\tilde{H}_{R} | \tilde{0} \rangle_R= E_{0R}=0.
\]
It turns out that the critical point is def\/ined by
\begin{gather}\label{eq19}
\sum_{i\neq j } \nu_{ij}= - (N-1)
\end{gather}
and tends to the result $\nu=-\frac{1}{N}$ in the limit of
identical particles. For $\nu_{ij}$ negative but greater than the
critical value \eqref{eq19}, the wave functions of the Hamiltonian
\eqref{eq14} are singular at coincidence points.

We conclude this Section with a few comments on the  multispecies
model in arbitrary dimen\-sions~\cite{Meljanac2}. Exactly in the
same way as was done in the one-dimensional case, we can f\/ind an
inf\/inite tower of exact eigenstates for $D$-dimensional
multispecies model. These eigenstates are given by $D$-dimensional
generalizations of the operators $\{ A_{1}^{\pm}, A_{2}^{\pm},
B_{2}^{\pm}\}$ and describe the part of the complete spectrum
which corresponds to center-of-mass states and global dilatation
states. The $D$-dimensional analogue of the critical point
\eqref{eq19} is simply $\sum\limits_{i\neq j } \nu_{ij}= -
(N-1)D$.

\section{Conclusion}
In this brief review we described some oscillator aspects of the
single-species and multispecies Calogero models in one dimension.
We have shown how both type of models could be obtained from the
most general (matrix) oscillator Hamiltonian by a suitable
``reduction'' procedure. We focused on the multispecies variant
and tried to solve it. Motivated by the successful operator
solution of the single-species Calogero model (based on the
$S_N$-extended Heisenberg algebra), we applied essentially the
same technique to its multispecies counterpart. It turned out that
we managed to reproduce only the f\/irst two Calogero-like modes.
We tried to evade this obstacle by considering underlying SU(1,1)
structure of the multispecies Hamiltonian. We def\/ined a new set
of creation and annihilation operators using the generators of
SU(1,1). Acting on the ground state, we were able to f\/ind
algebraically a class of the exact eigenstates and eigenenergies,
corresponding to the relative motion and motion of the center of
mass. This, SU(1,1)-based, approach  can be easy generalized to
the more sophisticated models, like multispecies Calogero model in
$D$ dimensions~\cite{Meljanac2} or conformal and   $PT$-invariant
Hamiltonians described in~\cite{Meljanac7}. However, the  problem
of f\/inding other exact eigenstates of the
 multispecies Calogero Hamiltonian(s) still remains open. Nevertheless,
 we hope that our analysis sheds some light on such kind of models.

\subsection*{Acknowledgments}
One of the authors (M.M.) would like to thank organizers of the
Sixth International Conference ``Symmetry in Nonlinear
Mathematical Physics'' (June 20--26, 2005, Kyiv) for their
invitation and warm hospitality. The present  paper is a written
version of the talk delivered by M.M.\ at this conference. It  is
supported by  Ministry of Science, Education and Sport of the
Republic of Croatia (contracts Nos. 0119261 and 0098003).

\LastPageEnding

\end{document}